\documentclass[preprint]{vldb}

\usepackage{balance}
\usepackage{enumitem}
\usepackage{graphicx}
\usepackage{hyperref}
\usepackage{listings}
\usepackage[labelformat=simple]{subcaption}
\usepackage{xcolor}

\lstdefinestyle{simd}{
  language=C,
  frame=single,
  basicstyle=\ttfamily\scriptsize,
  commentstyle=\color{blue},
  classoffset=2,
  morekeywords={const,int,__m512i},
  keywordstyle=\bfseries,
  classoffset=1,
  morekeywords={_mm512_slli_si512,_mm512_alignr_epi32,_mm512_add_epi32,_mm512_setzero_si512},
  keywordstyle=\color{red},
  classoffset=0,
}

\begin{document}

\title{Parallel Prefix Sum with SIMD}

\numberofauthors{3}
\author{
  \alignauthor
  Wangda Zhang\\
    \affaddr{Columbia University}\\
    \email{zwd@cs.columbia.edu}
  \alignauthor
  Yanbin Wang\\
    \affaddr{Columbia University}\\
    \email{yw3372@columbia.edu}
  \alignauthor
  Kenneth A. Ross\titlenote{This research was supported in part by a gift from Oracle Corporation.}\\
    \affaddr{Columbia University}\\
    \email{kar@cs.columbia.edu}
}
\date{21 June 2020}

\toappear{This article is published under a Creative Commons Attribution License (http://creativecommons.org/licenses/by/3.0/), which permits distribution and reproduction in any medium as well allowing derivative works, provided that you attribute the original work to the author(s) and ADMS 2020. \confname{11th International Workshop on Accelerating Analytics and Data Management Systems (ADMS’20), August 31, 2020, Tokyo, Japan.}}

\maketitle

\begin{abstract}
The prefix sum operation is a useful primitive with a broad range of applications. For database systems, it is a building block of many important operators including join, sort and filter queries. In this paper, we study different methods of computing prefix sums with SIMD instructions and multiple threads. For SIMD, we implement and compare horizontal and vertical computations, as well as a theoretically work-efficient balanced tree version using gather/scatter instructions. With multithreading, the memory bandwidth can become the bottleneck of prefix sum computations. We propose a new method that partitions data into cache-sized smaller partitions to achieve better data locality and reduce bandwidth demands from RAM. We also investigate four different ways of organizing the computation sub-procedures, which have different performance and usability characteristics. In the experiments we find that the most efficient prefix sum computation using our partitioning technique is up to 3x faster than two standard library implementations that already use SIMD and multithreading.
\end{abstract}

\section{Introduction}
\label{sec:intro}
\vspace{1.5em}

Prefix sums are widely used in parallel and distributed database systems as building blocks for important database operators. For example, a common use case is to determine the new offsets of data items during a partitioning step, where prefix sums are computed from a previously constructed histogram or bitmap, and then used as the new index values~\cite{blellochprefix}. Applications of this usage include radix sort on CPUs~\cite{satish2010fast,polychroniou2015rethinking} and GPUs~\cite{satish2009designing,leischner2010gpu}, radix hash joins~\cite{kim2009sort, barthels2017distributed}, as well as parallel filtering~\cite{sengupta2006work,billeter2009efficient}. In OLAP data cubes, prefix sums are precomputed so that they can be used to answer range sum queries at runtime~\cite{ho1997range,geffner1999relative,lemire2002wavelet}. Data mining algorithms such as K-means can be accelerated using parallel prefix sum computations in a preprocessing step~\cite{kohlhoff2012k}. In data compression using differential encoding, prefix sums are also used to reconstruct the original data, and prefix-sum computations can account for the majority of the running time~\cite{lemire2015decoding,lemire2016simd}.

The prefix sum operation takes a binary associative operator $\oplus$ and an input array of $n$ elements $[a_0,a_1,\dots,a_{n-1}]$, and outputs the array containing the sums of prefixes of the input: $[a_0,(a_0 \oplus a_1),\dots,(a_0 \oplus a_1 \oplus a_{n-1})]$. This definition is also known as an inclusive scan operation~\cite{blelloch1990vector}. The output of an exclusive scan (also called pre-scan) would remove the last element from the above output array, and insert an identity value at the beginning of the output. In this paper, we use addition as the binary operator $\oplus$ and compute inclusive scans by default.

The basic algorithm to compute prefix sums simply requires one sequential pass of additions while looping over all elements in the input array and writing out the running totals to the output array. Although this operation seems to be inherently sequential, there are actually parallel algorithms to compute prefix sums. To compute the prefix sums of $n$ elements, Hillis and Steele presented a data parallel algorithm that takes $O(\log n)$ time, assuming there are $n$ processors available~\cite{hillis1986data}. This algorithm performs $O(n\log n)$ additions, doing more work than the sequential version, which only needs $O(n)$ additions. Work-efficient algorithms~\cite{ladner1980parallel,blelloch1990vector} build a conceptual balanced binary tree to compute the prefix sums in two sweeps over the tree, using $O(n)$ operations. In Section~\ref{sec:simd}, we implement and compare SIMD versions of these data-parallel algorithms, as well as a vertical SIMD algorithm that is also work-efficient.

In a shared-memory environment, we can speed up prefix sum computations using multiple threads on a multicore platform. Prefix sums can be computed locally within each thread, but because of the sequential dependencies, thread $t_m$ has to know the previous sums computed by threads $t_0 \dots t_{m-1}$ in order to compute the global prefix sum results. Thus, a two-pass algorithm is necessary for multithreaded execution~\cite{singler2007mcstl}. There are multiple ways to organize the computation subprocedures, depending on (a) whether prefix sums are computed in the first or the second pass, and (b) how the work is partitioned. For example, to balance the work among threads, it is necessary to tune a dilation factor to the local configuration for optimal performance~\cite{singler2007mcstl}. To understand the best multithreading strategy, we analyze and compare different multithreaded execution strategies (Section~\ref{sec:thread:twopass}).

More importantly, with many concurrent threads accessing memory, the prefix sum can become a memory bandwidth-bound computation. To fully exploit the potential of the hardware, we must take care to minimize memory accesses and improve cache behavior. To this end, we propose to partition data into cache-sized partitions so that during a two-pass execution, the second pass can maximize its usage of the cache instead of accessing memory again (Section~\ref{sec:thread:partition}). We also develop a low-overhead thread synchronization technique, since partitioning into smaller data fragments can potentially increase the synchronization overhead.
%  among the threads.
% Our experiments show that cache-friendly partitioning can achieve significantly better performance than other external baselines.

In summary, the main contributions of this work are:
\begin{itemize}
\item We study multithreaded prefix sum computations, and propose a novel algorithm that splits data into cache-sized partitions to achieve better locality and reduce memory bandwidth usage.
\item We discuss data-parallel algorithms for implementing the prefix sum computation using SIMD instructions, in 3 different versions (horizontal, vertical, and tree).
\item We experimentally compare our implementations with external libraries. We also provde a set of recommendations for choosing the right algorithm.
\end{itemize}

\subsection{Related Work}
\vspace{1.5em}

Beyond databases, there are also many other uses of prefix sums in parallel computations and applications, including but not limited to various sorting algorithms (e.g., quicksort, mergesort, radix-sort), list ranking, stream compaction, polynomial evaluation, sparse matrix-vector multiplication, tridiagonal matrix solvers, lexical analysis, fluid simulation, and building data structures (graphs, trees, etc.) in parallel~\cite{blellochprefix,blelloch1990vector,sengupta2007scan}.

Algorithms for parallel prefix sums have been studied early in the design of binary adders~\cite{ladner1980parallel}, and analyzed extensively in theoretical studies~\cite{cole1989faster,chaudhuri1992complexity}. More recently, message passing algorithms were proposed for distributed memory platforms~\cite{sanders2006parallel}. A sequence of approximate prefix sums can be computed faster in $O(\log\log n)$ time~\cite{goldberg1995optimal}. For data structures, Fenwick trees can be used for efficient prefix sum computations with element updates~\cite{fenwick1994new}. Succinct indexable dictionaries have also been proposed to represent prefix sums compactly~\cite{raman2007succinct}.

As an important primitive, the prefix sum operation has been implemented in multiple libraries and platforms. The C++ standard library provides the prefix sum (scan) in its algorithm library, and parallel implementations are provided by GNU Parallel library~\cite{singler2008gnu} and Intel Parallel STL~\cite{intellibrary}. Parallel prefix sums can be implemented efficiently on GPUs with CUDA~\cite{harris2007parallel}, and for the Message Passing Interface (MPI-Scan)~\cite{sanders2006parallel}. Although compilers can not typically auto-vectorize a loop of prefix sum computation because of data dependency issues~\cite{maleki2011evaluation}, it is now possible to use OpenMP SIMD directives to dictate compilers to vectorize such loops. Since additions over floating point numbers are not associative, the result of parallel prefix sums of floating point values can have a small difference from the sequential computation.

\section{Thread-Level Parallelism}
\label{sec:thread}
\vspace{1.5em}

We now describe multithreaded algorithms for computing prefix sums in parallel. Note that a prefix-sum implementation can be either in-place, where the prefix sums replace the input data elements, or out-of-place, where the prefix sums are written to a new output array. The following description assumes an in-place algorithm. Out-of-place versions can be implemented similarly and will
be discussed in the experimental evaluation.
%\todo{discuss performance implication of out-of-place}

\subsection{Two-Pass Algorithms}
\label{sec:thread:twopass}
\vspace{1.5em}

\begin{figure}%[h]
  \centering
  \begin{subfigure}{0.5\textwidth}
    \centering
    \includegraphics[width=\textwidth]{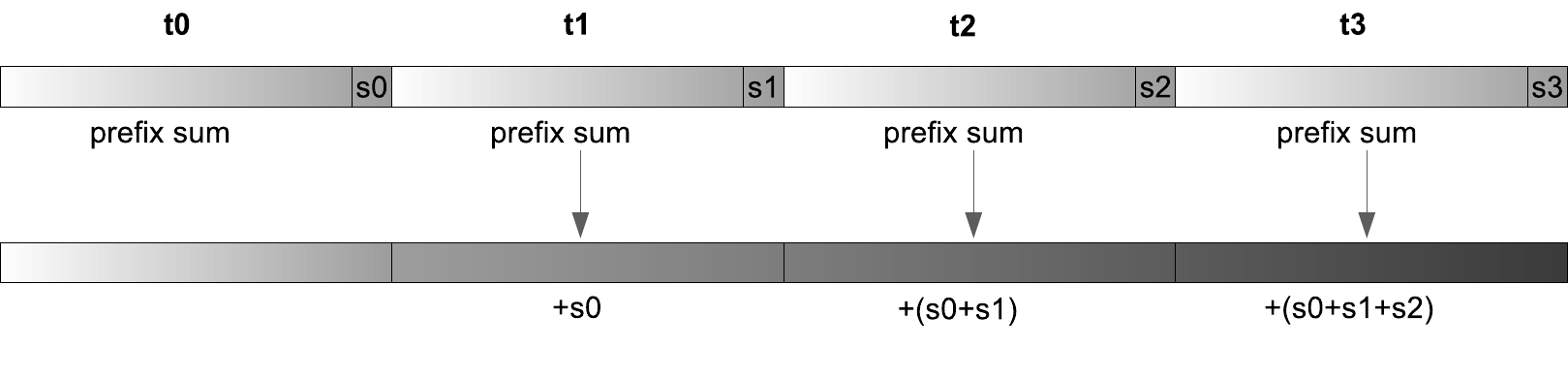}
    \caption{Prefix Sum + Increment}
    \label{fig:multithread1}
  \end{subfigure}
  \begin{subfigure}{0.5\textwidth}
    \centering
    \includegraphics[width=\textwidth]{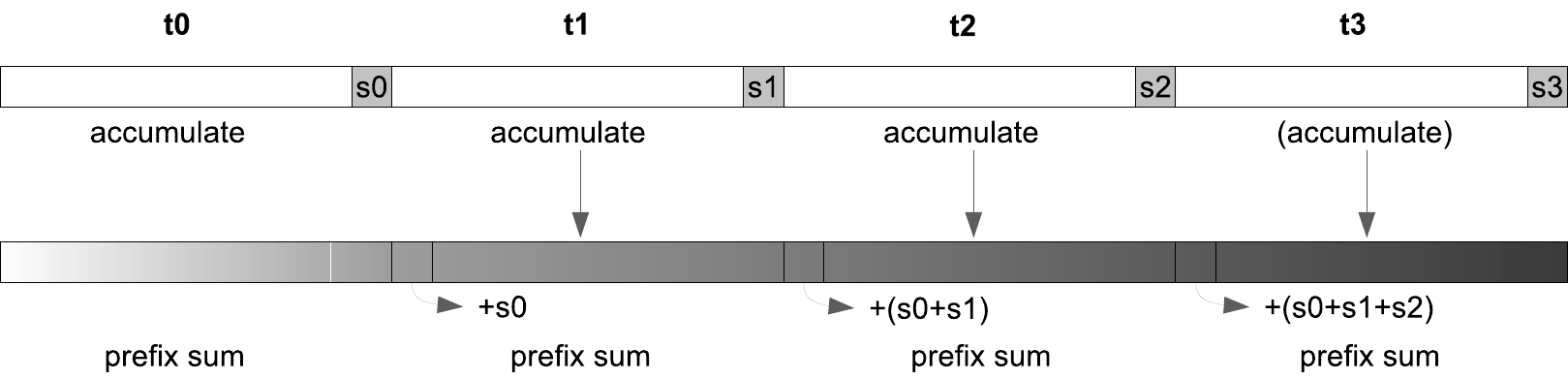}
    \caption{Accumulate + Prefix Sum}
    \label{fig:multithread2}
  \end{subfigure}
  \begin{subfigure}{0.5\textwidth}
    \centering
    \includegraphics[width=\textwidth]{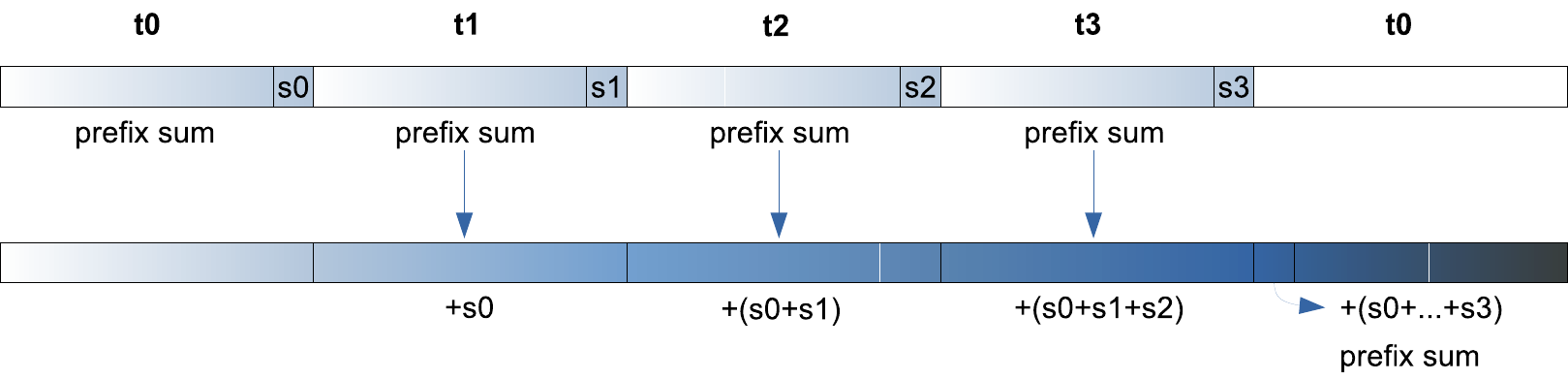}
    \caption{Prefix Sum + Increment (+1 partition)}
    \label{fig:multithread3}
  \end{subfigure}
  \begin{subfigure}{0.5\textwidth}
    \centering
    \includegraphics[width=\textwidth]{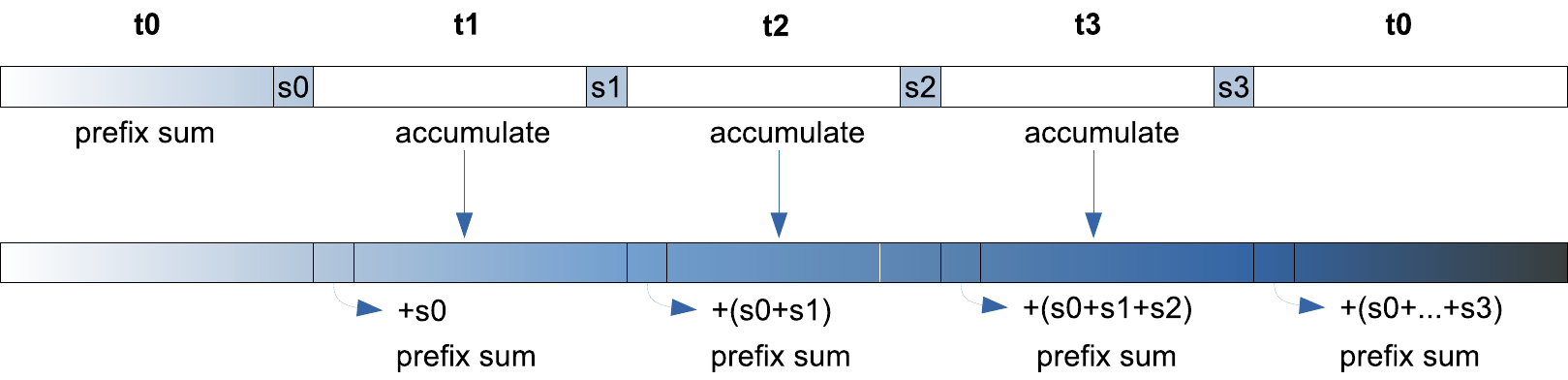}
    \caption{Accumulate + Prefix Sum (+1 partition)}
    \label{fig:multithread4}
  \end{subfigure}
  \caption{Multithreaded two-pass algorithms}
  \label{fig:multithread}
\end{figure}

To compute prefix sums in parallel, we study four different two-pass algorithms. There are two ways of processing the two passes depending on whether prefix sums are comptuted in the first or second pass. Figure~\ref{fig:multithread1} presents one implementation, using 4 threads ($t_0 \dots t_3$) as an example. To enable parallel processing, the data is divided into 4 equal-sized partitions. In the first pass, each thread computes a local prefix sum over its partition of data. After the prefix sum computation, the last element is the total sum of the partition, which will be used in the second pass to compute global prefix sums. These sums can be stored in a temporary buffer array $sums=\{s_0,s_1,s_2,s_3\}$, so it is easier to compute the prefix sums of their own. In the second pass, except for the first thread $t_0$, each thread $t_m$ increments every element of the local prefix sum by $\sum_{i<m} s_i$, in order to obtain global prefix sums. The prefix sums of the first partition are already the final results, since no increment is needed.

A potential inefficiency of this algorithm is that in the second pass, the first thread $t0$ is idle, so effectively the parallelism is only $(m-1)$ using $m$ threads, which has an impact on performance especially when $m$ is small. To fix this issue, the data can be divided into $(m+1)$ partitions, as shown in Figure~\ref{fig:multithread3} when $m=4$. In the first pass, the threads only work on the first 4 partitions, and the prefix sums of the last partition is computed in the second pass. We schedule thread $t0$ to work on the last partition, instead of shifting every thread to the next partition, since this is important for our partitioning technique in Section~\ref{sec:thread:partition}.

Figure~\ref{fig:multithread2} demonstrates a different way of computing the prefix sums. In the first pass, only the total sum of each partition is accumulated in parallel. Then in the second pass, the global prefix sums can be directly computed in parallel, using $\sum_{i<m} s_i$ as an offset to the input. The benefit of computing only the total sum in the first pass is that there is no memory write as in prefix sum computations. Therefore, this method can potentially require less memory bandwidth when the data size is large. In addition, we do not need the total sum of the last partition, so the last thread $t3$ can be idle in the first pass.

Similarly, Figure~\ref{fig:multithread4} fixes the idle-thread inefficiency by using one more partition. In the first pass, the prefix sums of $t0$ and the totals sums of $t1 \dots t3$ are computed in parallel. In the second pass, all threads compute prefix sums with an input offset computed from the $sums$ array. Thread $t0$ again works on the last partition. Both partitioning schemes in Figures~\ref{fig:multithread3} and \ref{fig:multithread4} ensure there are no idle threads and all threads work in parallel in either pass.

\subsubsection{Load Balancing}
\vspace{1.5em}

For algorithms shown in Figures~\ref{fig:multithread1} and \ref{fig:multithread2}, equal partitioning of the data should suffice since each thread does the same work (except for the idle thread). For Figure~\ref{fig:multithread3}, in the second pass, thread $t0$ computes prefix sums while other threads simply do an increment. Although in scalar code, both subprocedures require read, add and write, the increment is easily vectorizable by compilers while the prefix sum cannot be automatically vectorized. (We will also implement explicit SIMD versions of these algorithms.) Similarly, in the first pass of Figure~\ref{fig:multithread4}, thread $t0$ computes a prefix sum while other threads accumulate total sums. These operations may proceed at different rates, and the difference is potentially magnified because prefix sums cannot be autovectorized and need to write back results, while accumulation can be autovectorized and does not write to memory. (The second pass has similar problems when cache-friendly partitioning is used as we shall explain in Section~\ref{sec:thread:partition}.)

To compensate for thread $t0$ possibly taking longer, a dilation factor $d$ can be used to reduce the size of the first (or last) partition, in order to balance the work done by every thread~\cite{singler2007mcstl}. The range of a dilation factor is $d\in[0, 1]$, indicating the ratio of sizes between partition of thread $t0$ to other threads. When $d=0$, the corresponding partition is nonexistent, so Figure~\ref{fig:multithread1} is a special case of Figure~\ref{fig:multithread3} and Figure~\ref{fig:multithread2} is a special case of Figure~\ref{fig:multithread4}. When $d=1$, the partitions have equal sizes. In our experiments we find that the dilation factors have to be carefully tuned to achieve best performances, but in practice, standard library implementations typically just use equal-sized partitions by default, which is suboptimal in most cases. In fact, a poorly chosen dilation factor can cause an inefficiency worse than one idle thread, because many threads may become idle as they wait for thread t0 to complete its work.

\subsection{Cache-Friendly Partitioning}
\label{sec:thread:partition}
\vspace{1.5em}

For big data, one problem with the two-pass algorithms is that after the first pass, most data will have been evicted from the cache, and they have to be accessed again from memory in the second pass. Since the performance difference between cache and memory accesses is quite large, this problem can lead to huge performance impact overall. We therefore propose to partition the entire data in a cache friendly way, so that for each thread, the data it processes reside in cache after the first pass.

\begin{figure}
  \centering
  \includegraphics[width=0.5\textwidth]{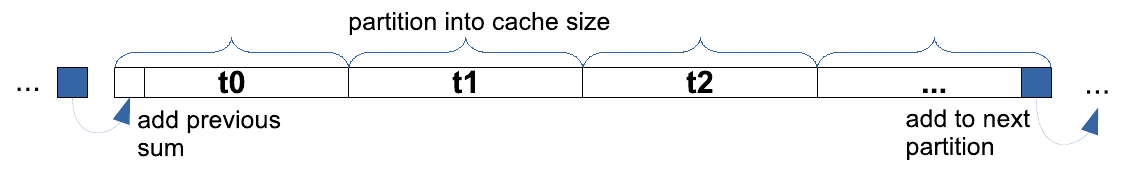}
  \caption{Cache-friendly partitioning}
  \label{fig:partition}
\end{figure}

Figure~\ref{fig:partition} demonstrates this idea, showing that the data are partitioned into cache-sized partitions, and processed by multiple threads in parallel. {\em Both} passes over the cache resident data are made (in parallel) before proceeding to the next partition. In this way, the second pass reads data from cache rather than from RAM. The entire data is processed sequentially in multiple iterations, but each iteration is computed in parallel by multiple threads. To compute global prefix sums, the first thread $t0$ needs to use the total sum from the previous partition (in the previous iteration) as an offset in the first pass.

If the partitioning uses the methods in Figures~\ref{fig:multithread1} or \ref{fig:multithread2}, then the first thread can use the last element of the previous partition as the offset, but this requires the first thread to wait for the second pass of the last partition to complete. A different way is to use the stored $sums$ array and compute the sum of all of its elements, including the local sum of last partition, as the offset. This way requires the local sum of the last partition (in the previous iteration) to have been computed, even if it is not necessary to do so using Accumulate in Figure~\ref{fig:multithread2}. However, there are two benefits of doing so: first, the data will reside in cache in the second pass, and second, we can save one synchronization after the second pass as we shall explain shortly. If the partitioning is done as in Figures~\ref{fig:multithread3} or \ref{fig:multithread4} using one more partition, then in the first pass, thread $t0$ can directly access the last element of its previous partition since the same thread $t0$ has processed it.

The effect of partitioning is that during the second pass, accesses to the same data can be served from the cache instead of memory. As we shall demonstrate in the experiments, the size of a partition is better set to roughly half of the size of L2 cache (or L1 cache if SIMD gather/scatter instructions are used in computing local prefix sums). As mentioned earlier, partitioning also changes the dilation factors used in the partitioning scheme in Figures~\ref{fig:multithread3} and \ref{fig:multithread4}, because the prefix sum computation in the second pass of the previous iteration will read from cache instead of memory, except for the last partition. For Figure~\ref{fig:multithread4}, we need a second dilation factor for the last partition as well to account for the difference.

\subsubsection{Thread Scheduling and Synchronization}
\vspace{1.5em}

To ensure the private cache of a physical core can be accessed during the second pass, we control the thread affinity so that physical threads $t1, t2, \dots$ continue to process the same data in the second pass. Thread $t0$ works on a different partition, since the first partition does not need a second pass while the last partition does not need the first pass.

We need a barrier synchronization after the first pass, and all threads should wait for it before continuing to the second pass. Without the cache-friendly partitioning, the original two-pass algorithm only needs to synchronize twice: one between the first pass and the second pass, and one to join all the threads after the second pass. With partitioning, more synchronization points are needed for this multi-iteration process. However, we only need one synchronization in every iteration, instead of two. The synchronization after the second pass is unnecessary, and the second pass of iteration $k$ can be overlapped with the first pass of iteration $(k+1)$. In order to prevent the local sums in the $sums$ array of iteration $k$ from being overwritten by sums computed in the first pass of iteration $(k+1)$, we need two $sums$ arrays to store the local sums. In iteration $k$, the sums are stored in $sums_{k\%2}$. Because of the synchronization between the two passes, this double buffering of $sums$ is sufficient.

Since we still need one synchronization in every iteration, the overhead of our synchronization mechanism should be as low as possible. As reported by previous studies~\cite{hetland2019paths,rodchenko2015effective}, the latency of different software barrier implementations can differ by orders of magnitude, thus profoundly affecting the overall performance. In our implementation we use a hand-tuned reusable counting barrier, which is implemented as a spinlock with atomic operations. On our experimental platforms with 48 cores, we find its performance is much better than a Pthread barrier and slightly better than the OpenMP barrier. To scale to even more cores, it is potentially beneficial to use even faster barrier implemenations (e.g., tournament barrier~\cite{hetland2019paths}). Because of the sequential dependency of prefix sum computations, it is also possible to restrict synchronization to only adjacent threads, rather than using a centralized barrier~\cite{yan2013streamscan}.
% We also observe that for prefix sum, thread $t_m$ just needs to wait for its previous threads $t_i, i<m$ to finish. Therefore, instead of use one centralized barrier, we can use a chain of barriers so that thread $t_m$ only waits for its previous thread $t_{m-1}$. Once the barrier is cleared, it then notifies the next thread to continue.

% NUMA

\section{Data-Level Parallelism}
\label{sec:simd}
\vspace{1.5em}

SIMD instructions can be used to speed up the single-thread prefix sum computation. In this paper, we use AVX-512 extensions that have more parallelism than previous extensions. We assume 16 32-bit numbers can be processed in parallel using the 512-bit vectors.

\subsection{Horizontal}
\label{sec:simd:h}
\vspace{1.5em}

\begin{figure}
  \centering
  \includegraphics[width=0.3\textwidth]{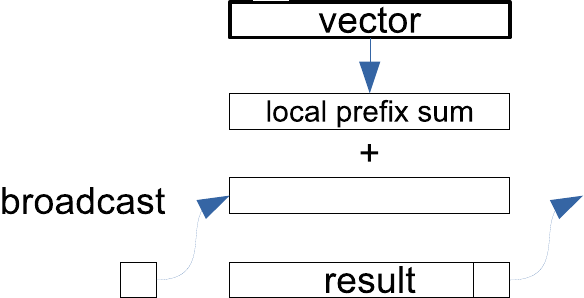}
  \caption{Horizontal SIMD}
  \label{fig:simd:h}
\end{figure}

% \todo{bold fonts not working}
\begin{lstlisting}[float,style=simd,caption={In-register prefix sums with Horizontal SIMD},captionpos=b,label={lst:simd:h}]
__m512i _mm512_slli_si512(__m512i x, int k) {
  const __m512i ZERO = _mm512_setzero_si512();
  return _mm512_alignr_epi32(x, ZERO, 16 - k);
}
__m512i PrefixSum(__m512i x) {
  x = _mm512_add_epi32(_mm512_slli_si512(x, 1)));
  x = _mm512_add_epi32(_mm512_slli_si512(x, 2)));
  x = _mm512_add_epi32(_mm512_slli_si512(x, 4)));
  x = _mm512_add_epi32(_mm512_slli_si512(x, 8)));
  return x; // local prefix sums
}
\end{lstlisting}

Figure~\ref{fig:simd:h} shows how we can compute prefix sums in a horizontal way using SIMD instructions. The basic primitive is to compute the prefix sum of a vector of 16 elements in register, so that we can loop over the data to compute every 16 elements. The last element of the 16 prefix sums results is then broadcast into another vector, so it can be added to the next 16 elements to compute the global results.

As mentioned in Section~\ref{sec:intro}, \cite{hillis1986data} presented a data parallel algorithm to compute prefix sums, which can be used to compute the prefix sums in a register. For $w$ elements in a SIMD register, this algorithm uses $\log(w)$ steps. So for 16 four-byte elements, we can compute the prefix sums in-register using 4 shifts and 4 addition instructions. The pseudocode in Listing~\ref{lst:simd:h} shows this procedure.

Note that unlike previous SIMD extensions (e.g. AVX2), AVX-512 does not provide the {\tt \_mm512\_slli\_si512} instruction to shift the 512-bit vector. Here, we implement the shift using the {\tt valign} instruction. It is also possible to implement this shifting using permutate instructions~\cite{intellibrary,klemm2012extending}, but more registers have to be used to control the permutation destinations for different shifts, and it appears to be slightly slower (although {\tt valign} and {\tt vperm} use the same number of cycles).

% every 16 elements: 4 add + 4 shift + 1 broadcast

\subsection{Vertical}
\label{sec:simd:v}
\vspace{1.5em}

\begin{figure}%[ht!]
  \centering
  \includegraphics[width=0.3\textwidth]{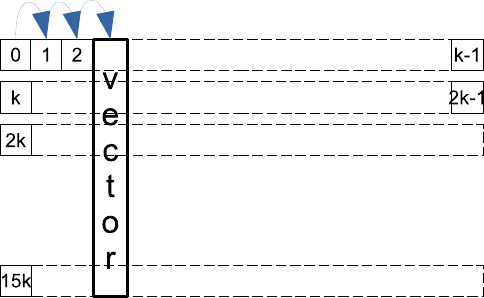}
  \caption{Vertical SIMD}
  \label{fig:simd:v}
\end{figure}

We can also compute the prefix sums vertically by dividing the data into 16 chunks of length $k=n/16$ and then using a vector to accumulate the running sums, as shown in Figure~\ref{fig:simd:v}. While scanning over the data, we gather the data at indices $(0+j, k+j, 2k+j, \dots, 15k+j)$ (where $j\in[0,k)$), add to the vector of running sums, and scatter the result back to the original indices position (assuming the indices fit into four bytes). After one pass of scanning over the data, SIMD lane $i$ contains the total sum of elements from $i*k$ to $(i+1)*k-1$. In a similar way to the multithreaded executions (Section~\ref{sec:thread}), we need to update the results in the second pass.

In addition, we can also switch the two passes by first computing just the total sum of each chunk (without the scatter step) in the first pass, and then compute the global prefix sums vertically in the second pass.

No matter which pass prefix sums are computed in, we need two passes for vertical SIMD computation because of data-parallel processing even if it is executed in a single thread. For multithreaded execution, each of the two passes can run in parallel among the threads. Similar to the cache-friendly partitioning for multithreaded execution, we can also partition the data into cache sizes so that the second pass of vertical SIMD implementation can reuse the data from cache (even for single thread).

Different from the horizontal SIMD version, this vertical algorithm is work-efficient, performing $O(n)$ additions (including updating the indices for gather/scatter) without the $O(\log w)$ overhead. Using data-parallel execution, we essentially execute a two-pass sequential algorithm for each of the 16 chunks of data. In practice, the $\log w$ extra additions in the horizontal version are computed in register, which is much faster than waiting for memory accesses. Even so, if the gather/scatter instructions are fast, then the vertical algorithm can potentially be faster than the horizontal version.

% every 16 elements: 1 gather + 1 add + 1 scatter + 1 add (+ 2 add)
% limitation: indices must fit into 32 bits for 16 parallel SIMD lanes.

\subsection{Tree}
\label{sec:simd:t}
\vspace{1.5em}

\begin{figure}%[b]
  \centering
  \includegraphics[width=0.45\textwidth]{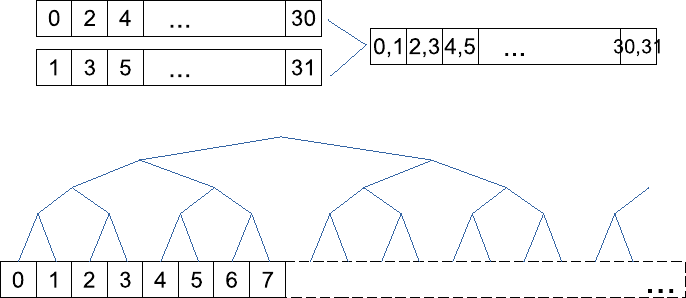}
  \caption{Tree SIMD}
  \label{fig:simd:t}
\end{figure}

As introduced in Section~\ref{sec:intro}, a work-efficient algorithm using two sweeps over a balanced tree is presented in~\cite{blellochprefix}. As a two-pass algorithm, the first pass performs an up-sweep (reduction) to build the tree, and the second pass performs a down-sweep to construct the prefix sum results. At each level of the tree, the computation can be done in a data parallel way for the two children of every internal tree node. Figure~\ref{fig:simd:t} shows the conceptual tree built from the data; for details of the algorithm, refer to the original paper.

To implement this algorithm, we can reuse the {\tt PrefixSum()} procedure in Section~\ref{sec:simd:h} and mask off the unnecessary lanes in the first pass; the second pass can be done similarly with reversed order of instructions. However, it does not make sense to artificially make the SIMD lanes idle. Instead, we can use a loop of gather/scatter instructions to process the elements in a strided access pattern at every level of the tree. Because the tree is of height $\log n$, the implementation needs multiple passes of gathers and scatters, so although the algorithm is work-efficient in term of addition, it is quite inefficient in memory access. Cache-friendly partitioning helps alleviate this issue for the second pass, but overall this algorithm is not suitable for SIMD implementations.

\section{Experiments}
\vspace{1.5em}

\subsection{Setup}
\vspace{1.5em}

We conducted experiments using the Amazon EC2 service on a dedicated m5.metal instance\footnote{\url{https://aws.amazon.com/intel}}, runing in non-virtualized environments without hypervisors. The instance has two Intel Xeon Platinum 8175M CPUs, based on the Skylake microarchitecture that supports AVX-512. Table~\ref{tab:hardware} describes the platform specifications. The instance provides 300 GB memory, and the measured memory bandwidth is 180 GB/s on two NUMA nodes in total.

\begin{table}
  \centering
  \caption{Hardware Specification}
  \label{tab:hardware}
  \begin{tabular}{l|c} \hline
    Microarchitecture & ~~~~Skylake~~~~ \\  \hline
    Model             & 8175M           \\  \hline
    Sockets           & 2               \\  \hline
    NUMA nodes        & 2               \\  \hline
    Cores per socket  & 24              \\  \hline
    Threads per core  & 2               \\  \hline
    Clock frequency   & 2.5 GHz         \\  \hline
    L1D cache         & 32 KB           \\  \hline
    L2 cache          & 1 MB            \\  \hline
    L3 cache (shared) & 33 MB           \\  \hline
 \end{tabular}
\end{table}

We implemented the methods in Sections~\ref{sec:simd} and \ref{sec:thread} using Intel AVX-512 intrinsics and POSIX Threads with atomic operations for synchronization. Our code\footnote{\url{http://www.cs.columbia.edu/~zwd/prefix-sum}} was written in C++, compiled using Intel compiler 19.1 with {\tt -O3} optimization (scalar code, such as subprocedures Increment and Accumulate, can be autovectorized), and ran on 64-bit Linux operating systems. For our experiemnts, we use 32-bit floating point numbers as the data type, and generate random data so that every thread has 128 MB of data (i.e. 32 million floats) to work with. Algorithms are in-place prefix sum computations, except in Section~\ref{sect-oop} where we consider out-of-place algorithms.
%\todo{out-of-place}

Our handwritten implementations used in the experiments are summarized in Table~\ref{tab:algo}. We compare with the following two external baselines:
\begin{itemize}
  \item GNU Library~\cite{gnulibrary}. The {\tt libstdc++} parallel mode provides a multithreaded prefix sum implementation using OpenMP. The implementation executes a two-pass scalar computation (Accumulate + Prefix Sum) as discussed in Section~\ref{sec:thread}.
  \item Intel Library~\cite{intellibrary}. Intel provides a Parallel STL library, which is an implementation of the C++ standard library algorithms with multithreading and vectorized execution (under the {\tt par\_unseq} execution policy). The multithreaded execution is enabled by Intel Threading Building Blocks (TBB), and we can force the Intel compiler to use 512-bit SIMD registers by using the {\tt -qopt-zmm-usage=high} flag (which is set to low on Skylake by default).
\end{itemize}

\begin{table*}
  \centering
  \caption{Algorithm Descriptions}
  \label{tab:algo}
  \begin{tabular}{l|l} \hline
    Scalar & Single-thread one-pass scalar implementation \\  \hline
    SIMD & Single-thread one-pass horizontal SIMD (Section~\ref{sec:simd:h}). Also used for multithread implementations \\ \hline
    SIMD-V1/V2 & Single-thread two-pass vertical SIMD (Section~\ref{sec:simd:v}), computing prefix sums in Pass 1 / Pass 2 \\ \hline
    SIMD-T & Single-thread two-pass tree SIMD (Section~\ref{sec:simd:t}) \\  \hline
    Scalar1, SIMD1 & Multithread two-pass algorithm, computing prefix sums in Pass 1 (Figure~\ref{fig:multithread3})\\  \hline
    Scalar2, SIMD2 & Multithread two-pass algorithm, computing prefix sums in Pass 2 (Figure~\ref{fig:multithread4})\\  \hline
    Scalar*-P, SIMD*-P & Multithread two-pass algorithm with cache-friendly partitioning (Section~\ref{sec:thread:partition})\\  \hline
 \end{tabular}
\end{table*}

\subsection{Results}
\vspace{1.5em}

% \begin{figure*}
%   \centering
%   \begin{minipage}{0.45\textwidth}
%     \centering
%     \includegraphics[width=0.9\linewidth]{figures/throughput_singlethread}
%     \captionof{figure}{Single-thread throughput}
%     \label{fig:exp:singlethread}
%   \end{minipage}%
%   \begin{minipage}{0.45\textwidth}
%     \centering
%     \includegraphics[width=0.9\linewidth]{figures/throughput_multithread}
%     \captionof{figure}{Multithread throughput}
%     \label{fig:exp:multithread}
%   \end{minipage}
% \end{figure*}

\subsubsection{Single-Thread Performance}
\vspace{1.5em}

\begin{figure}
  \centering
  \includegraphics[width=0.45\textwidth]{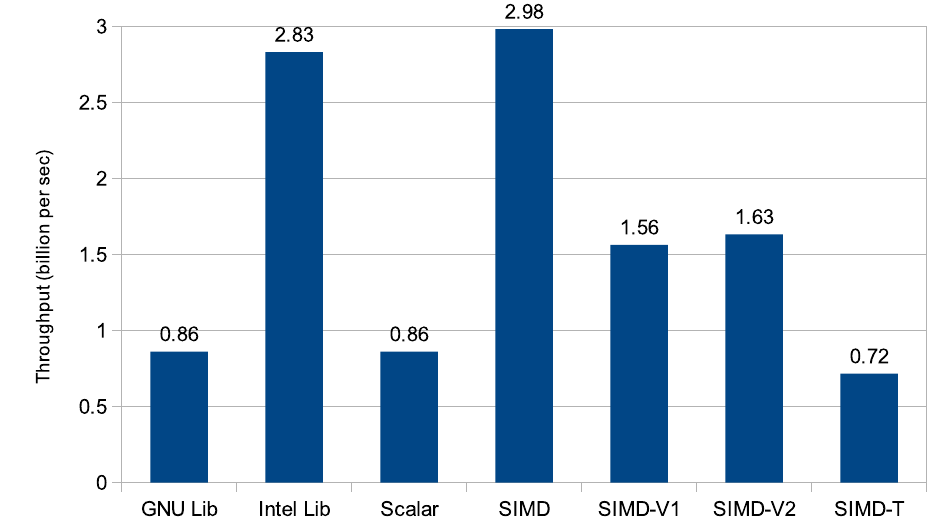}
  \caption{Single-thread throughput}
  \label{fig:exp:singlethread}
\end{figure}

Figure~\ref{fig:exp:singlethread} presents the single-thread throughput results. The (horizontal) SIMD implementation (Section~\ref{sec:simd:h}) using AVX-512 performs the best, processing nearly three billion floats per second.
The single-thread Intel library implementation with vectorized execution is slightly slower. Our scalar implementation has the same performance as the scalar GNU library implementation, which is 3.5x slower than SIMD.

Using cache-friendly partitioning (with the best partition sizes), the vertical SIMD implementations (Section~\ref{sec:simd:v}) are still about 2x slower than horizontal SIMD. Computing prefix sums in the second pass is slightly better. The Tree implementation (Section~\ref{sec:simd:t}) is slowest even with partitioning because of its non-sequential memory accesses.

\subsubsection{Multithread Performance}
\vspace{1.5em}

\begin{figure*}%[h]
  \centering
  \begin{subfigure}{0.45\textwidth}
    \centering
    \includegraphics[width=\textwidth]{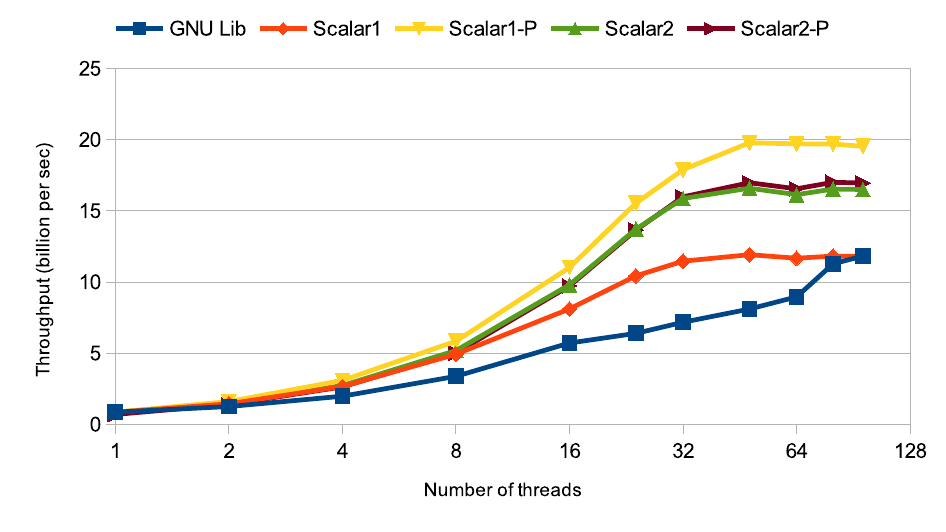}
    \caption{Scalar}
    \label{fig:exp:multithread:scalar}
  \end{subfigure}
  \begin{subfigure}{0.45\textwidth}
    \centering
    \includegraphics[width=\textwidth]{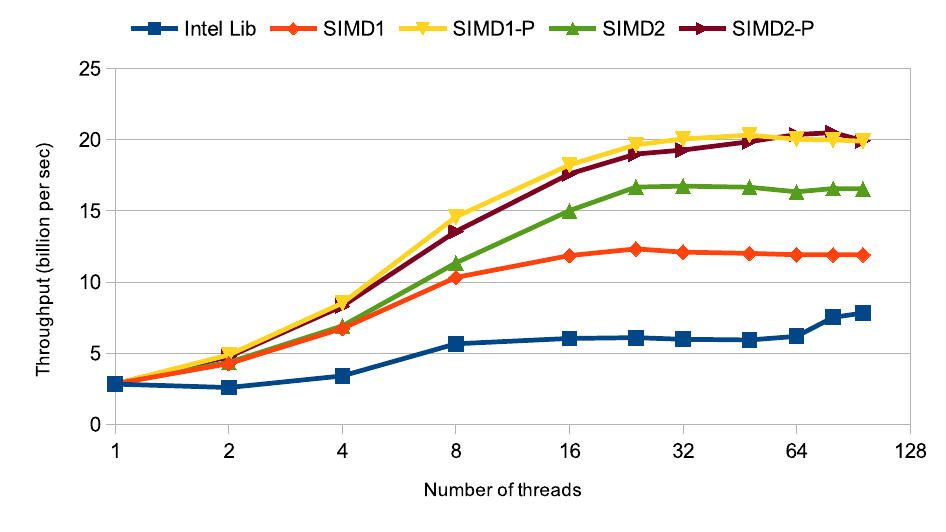}
    \caption{SIMD}
    \label{fig:exp:multithread:simd}
  \end{subfigure}
  \caption{Multithread throughput}
  \label{fig:exp:multithread}
\end{figure*}

We then increase the number of threads used to the maximum number of (hyper-)threads supported on the platform. In addition to external baselines, we compare our implementations with and without partititioning. The partition sizes and dilation factors used are the best ones as we shall discuss in Section~\ref{sec:exp:partition}. Since the vertical SIMD and tree implementations are slow, we omit them in multithreaded experiments.

Figure~\ref{fig:exp:multithread:simd} presents the multithreaded throughput results using SIMD. The partitioned SIMD implementations are consistently the best methods across all numbers of threads. SIMD1-P has a throughput of 20.3 billion per second at 48 threads. At this point, it already saturates the memory bandwidth, so its performance does not improve with more than 48 threads used. The highest throughput of SIMD2-P is 20.5 billion per second with 80 threads. Without partitioning, the SIMD1 stablizes at a throughput of 12.3 billion per second. Because the second pass has to access the data from memory (instead of cache) again without partitioning, it is 1.7x slower. SIMD2 has a throughput of 16.7 billion per second, 1.3x faster than SIMD1 because of less memory access in the first pass (no writes).

Figure~\ref{fig:exp:multithread:scalar} presents the scalar results. Interestingly, although the partitioned scalar implementation Scalar1-P is slow with less than 32 threads, it eventually becomes faster than the non-partitioning SIMD method and also reaches the limit of memory bandwidth at 48 threads. This result shows that the prefix sum converts from a CPU-bound computation to a memory-bound computation as more and more threads are used. In a compute-bound situation (e.g., less than 32 threads), SIMD can improve performance, while in a memory-bandwidth-bound situation, it is more important to optimize for cache locality and reduce memory access. For Scalar2-P, partitioning does not appear beneficial. We think one possible explanation would be that the compiler (or hardware prefetcher) generates nontemporal prefetches (or loads) in the first accumulation pass, meaning that the data is not resident in the cache for the second pass. The non-partitioning scalar implementations are slower than SIMD, but with more threads used, they also become bandwidth-bound and reach the same performance as non-partitioning SIMD.

We also find that library implementations are slower than our handwritten implementations. The Intel library implementation is faster than the GNU library with less than 16 threads, since it uses SIMD implementations, but it appears to not scale well. Even comparing with our non-partitioning implementations, both GNU and Intel implementations have a lower throughput at the maximum system capacity.

\subsubsection{Out-of-Place Performance}
\label{sect-oop}
\vspace{1.5em}

\begin{figure*}%[h]
  \centering
  \begin{subfigure}{0.45\textwidth}
    \centering
    \includegraphics[width=\textwidth]{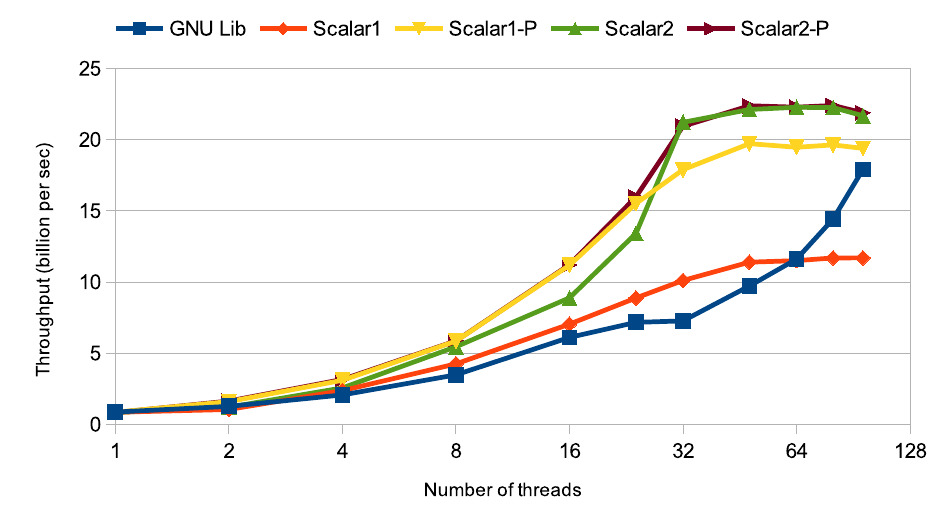}
    \caption{Scalar}
    \label{fig:exp:multithread:oop:scalar}
  \end{subfigure}
  \begin{subfigure}{0.45\textwidth}
    \centering
    \includegraphics[width=\textwidth]{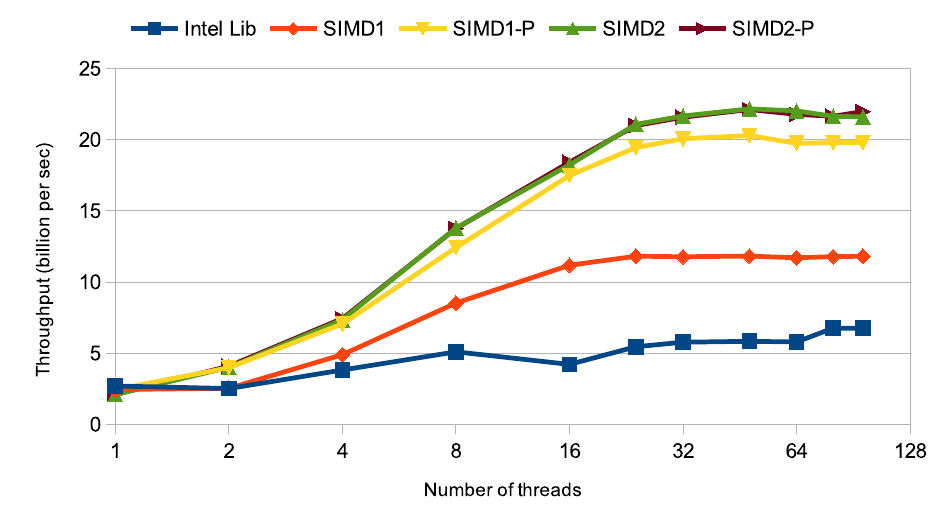}
    \caption{SIMD}
    \label{fig:exp:multithread:oop:simd}
  \end{subfigure}
  \caption{Multithread throughput (out-of-place)}
  \label{fig:exp:multithread:oop}
\end{figure*}

%\todo{out-of-place results}
We now investigate out-of-place computations, where the output goes to a separate array from the input.  Figure~\ref{fig:exp:multithread:oop} shows that Scalar2 and SIMD2 perform well; partitioned versions of those algorithms giving about the same throughput, except for scalar code with few threads where the partitioned algorithm performs better. The performance of the GNU library improves for out-of-place computations, but it still performs worse than the Scalar2/Scalar2-P algorithms.  While there is little change in the performance of the Scalar1/Scalar1-P and SIMD1/SIMD1-P algorithms when shifting from in-place to out-of-place computations, it appears that the performance of Scalar2/Scalar2-P and SIMD2/SIMD2-P improves.

\begin{figure*}%[h]
  \centering
  \begin{subfigure}{0.45\textwidth}
    \centering
    \includegraphics[width=\textwidth]{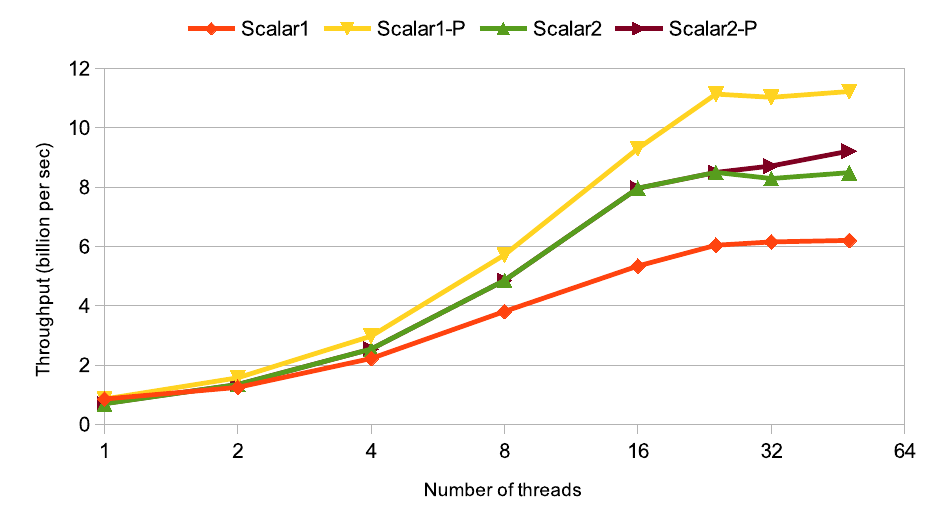}
    \caption{Scalar (in-place)}
    \label{fig:exp:onenode1}
  \end{subfigure}
  \begin{subfigure}{0.45\textwidth}
    \centering
    \includegraphics[width=\textwidth]{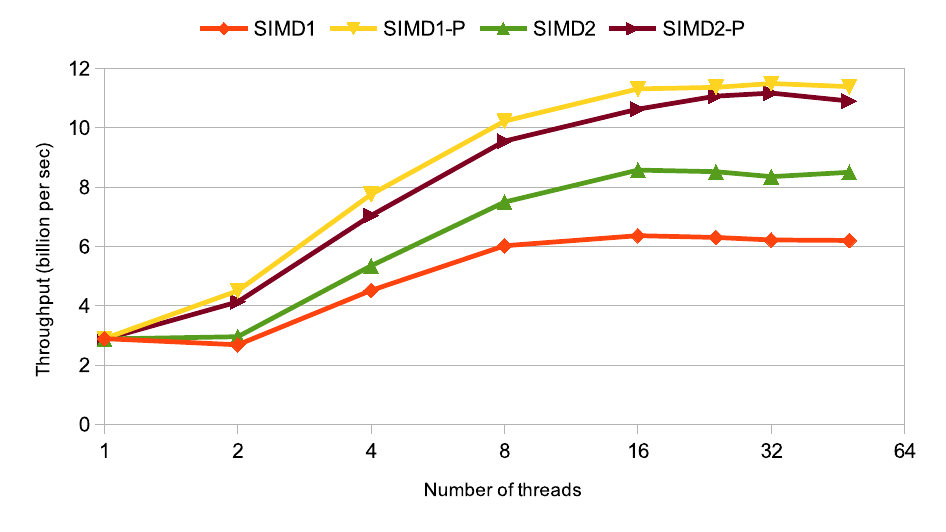}
    \caption{SIMD (in-place)}
    \label{fig:exp:onenode2}
  \end{subfigure}
  \begin{subfigure}{0.45\textwidth}
    \centering
    \includegraphics[width=\textwidth]{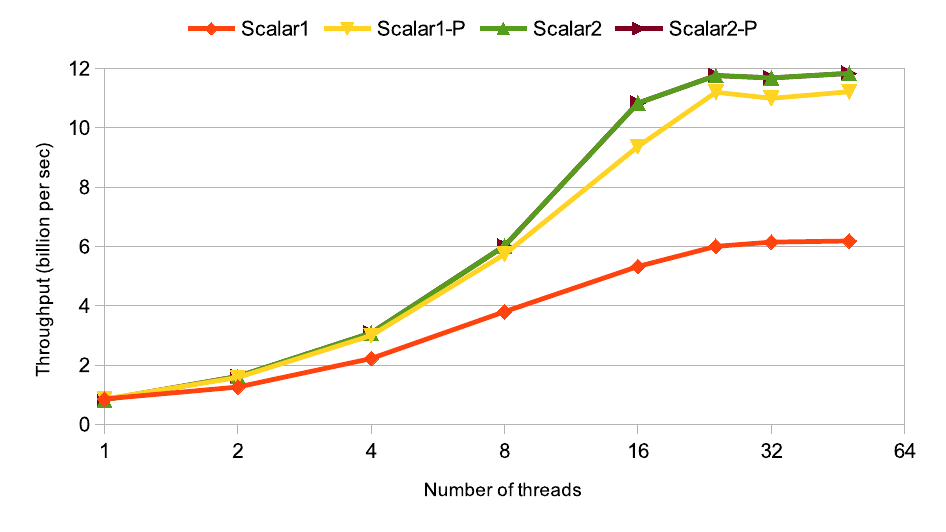}
    \caption{Scalar (out-of-place)}
    \label{fig:exp:onenode3}
  \end{subfigure}
  \begin{subfigure}{0.45\textwidth}
    \centering
    \includegraphics[width=\textwidth]{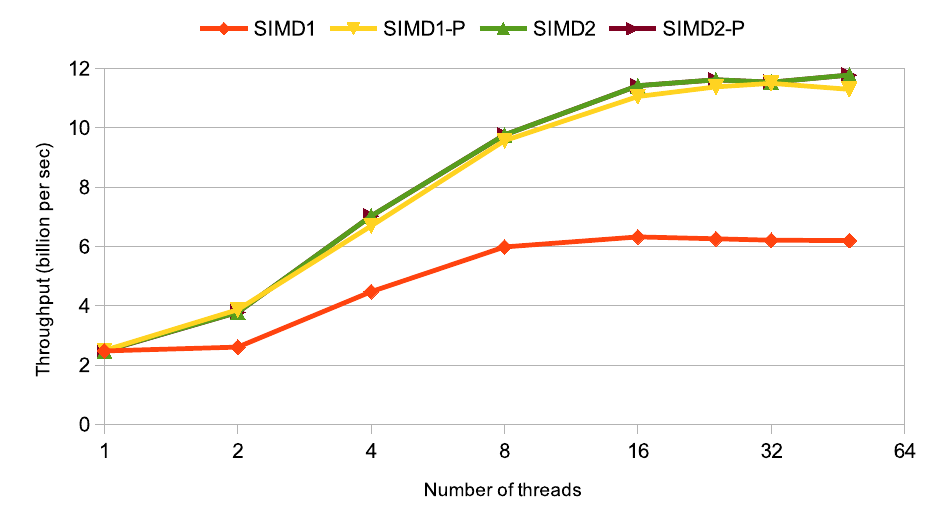}
    \caption{SIMD (out-of-place)}
    \label{fig:exp:onenode4}
  \end{subfigure}
  \caption{Throughput on a single node}
  \label{fig:exp:onenode}
\end{figure*}

One source of potential improvement is that the out-of-place method is reading from/writing to different memory regions, which may be on different memory banks that can operate in parallel. The in-place method will always be reading/writing to a single memory bank at a time.  There are therefore more opportunities for the system to balance the throughput from different memory banks, and achieve higher bandwidth utilization. Support for this hypothesis comes from Figure~\ref{fig:exp:onenode} where we run the algorithms on a single node with a single memory bank.  The SIMD versions of SIMD1-P, SIMD2 and  SIMD2-P now all perform similarly.

Nevertheless, the scalar performance in Figure~\ref{fig:exp:onenode} shows that there is still a small performance edge for Scalar2/Scalar2-P in the out-of-place algorithms on a single node. We are not certain of the explanation for this effect, but suspect that write-combining buffers may be helping the performance, since the second phase in Scalar2/Scalar2-P does sequential blind writes to the output array.

\subsubsection{Effect of Partition Sizes}
\label{sec:exp:partition}
\vspace{1.5em}

\begin{figure*}%[h]
  \centering
  \begin{subfigure}{0.45\textwidth}
    \centering
    \includegraphics[width=\textwidth]{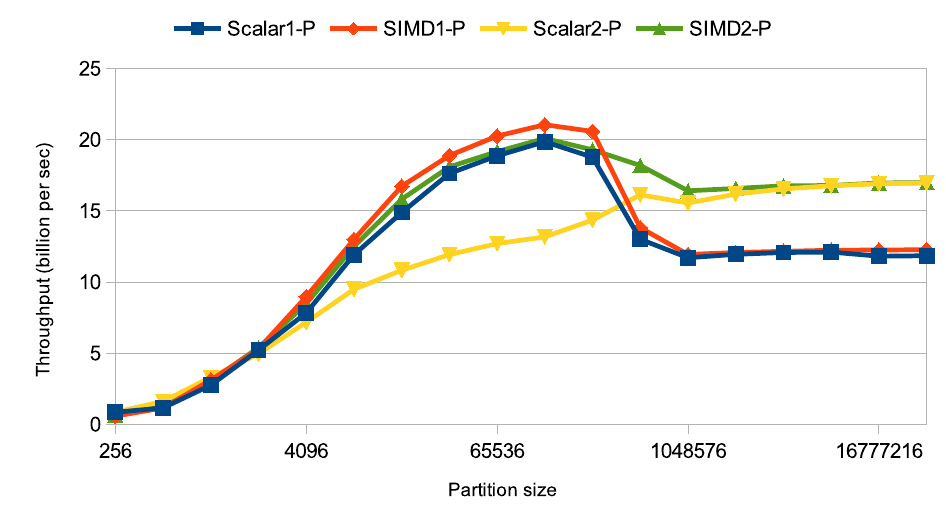}
    \caption{Multithreaded implementations with partitioning}
    \label{fig:exp:partition1}
  \end{subfigure}
  \begin{subfigure}{0.45\textwidth}
    \centering
    \includegraphics[width=\textwidth]{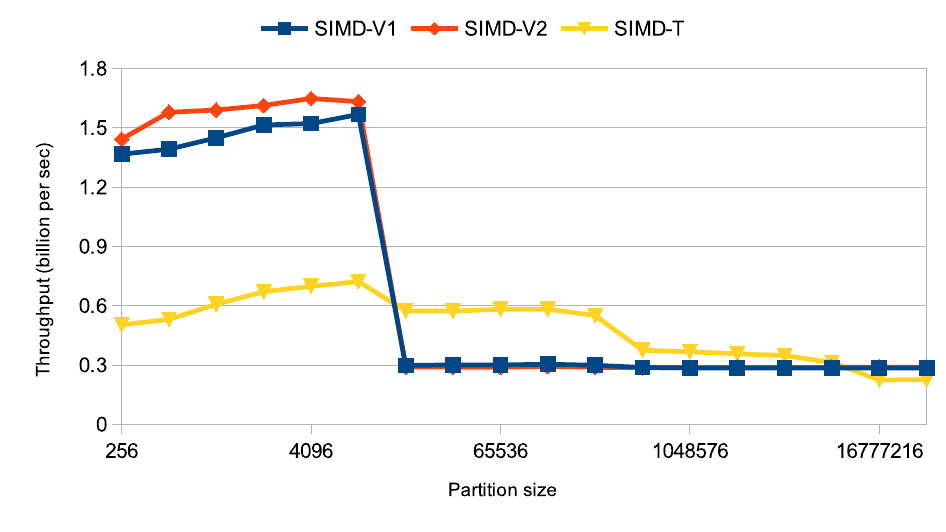}
    \caption{Single-thread implementations with partitioning}
    \label{fig:exp:partition2}
  \end{subfigure}
  \caption{Effect of partition sizes}
  \label{fig:exp:partition}
\end{figure*}

In Figure~\ref{fig:exp:partition1}, we tune the partition sizes for our partitioned scalar and SIMD implementations with 48 threads. The x-axis is in log scale. By trying out possible partition sizes covering the cache sizes at different levels, we find that with 48 (and fewer) threads, the best partition size is 128K floats per thread, which takes 512 KB, i.e. half the size of L2 caches. With 96 threads, the best choice of partition size is 64K floats per thread, as a result of two hyperthreads sharing the L2 cache. Large partition sizes make caching ineffective, while small partition sizes can increase the synchronization overhead.

Partitioning also helps with the vertical and tree implementations, which need two passes even with a single thread. In Figure~\ref{fig:exp:partition2}, we find that by partitioning the input data into partitions of 8K floats (L1 cache size), we can increase the throughput of SIMD-V1 and SIMD-V2 from 0.3 to about 1.6 billion per second, and the throughput of SIMD-T from 0.2 to 0.7 billion per second. This result shows that partitions fitting into L1 cache is best for the performance of gather and scatter instructions. Even with this enhancement, the throughput is still much lower than other algorithms.

\subsubsection{Effect of Dilation Factors}
\vspace{1.5em}

\begin{figure*}%[h]
  \centering
  \begin{subfigure}{0.45\textwidth}
    \centering
    \includegraphics[width=\textwidth]{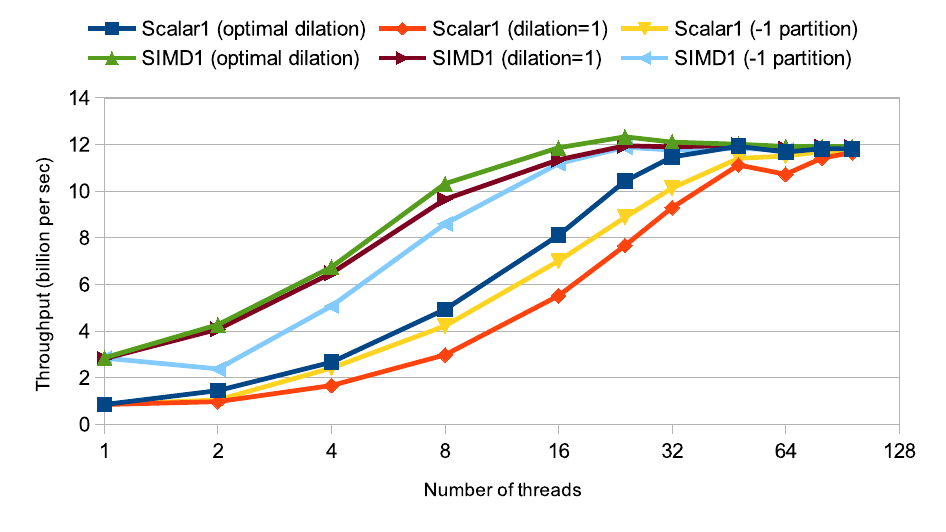}
    \caption{Prefix sum + Increment, no partitioning}
    \label{fig:exp:dilation1}
  \end{subfigure}
  \begin{subfigure}{0.45\textwidth}
    \centering
    \includegraphics[width=\textwidth]{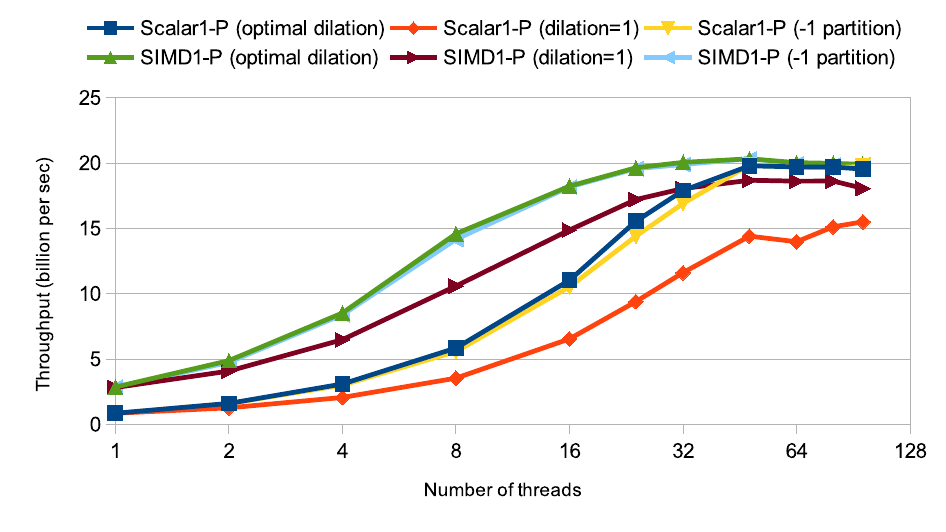}
    \caption{Prefix sum + Increment}
    \label{fig:exp:dilation2}
  \end{subfigure}
  \begin{subfigure}{0.45\textwidth}
    \centering
    \includegraphics[width=\textwidth]{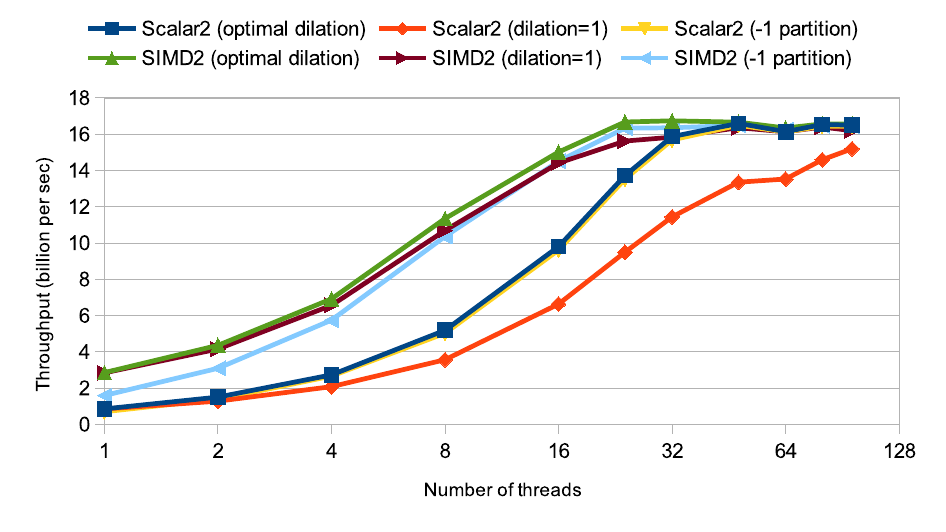}
    \caption{Accumulate + Prefix Sum, no partitioning}
    \label{fig:exp:dilation3}
  \end{subfigure}
  \begin{subfigure}{0.45\textwidth}
    \centering
    \includegraphics[width=\textwidth]{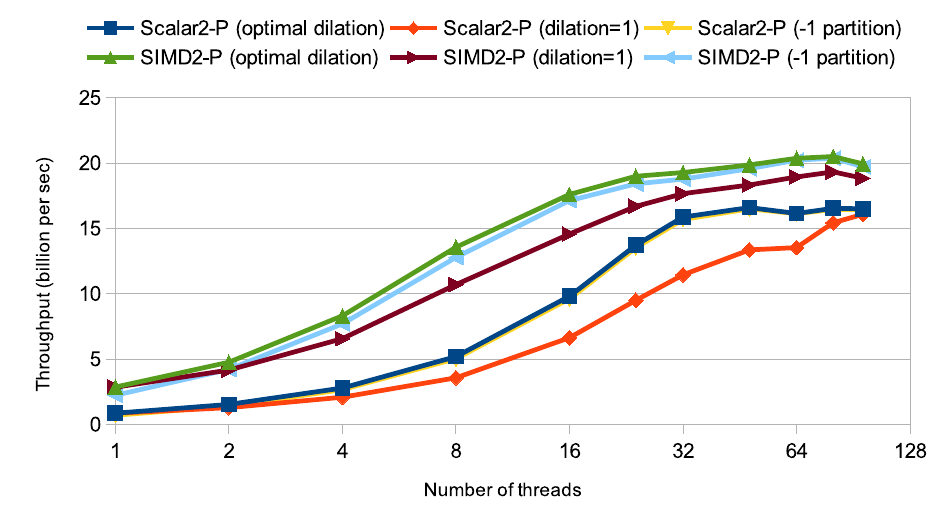}
    \caption{Accumulate + Prefix Sum}
    \label{fig:exp:dilation4}
  \end{subfigure}
  \caption{Effect of dilation factors}
  \label{fig:exp:dilation}
\end{figure*}

Figure~\ref{fig:exp:dilation} demonstrates the effect of dilation factors. We compare equal partitioning (no dilation) with best dilation factors found in our tuning process. In most situtations shown, it is clear that tuning the dilation factor is required to achieve good performance.

The improvement of using one more partition (in every iteration) is largest when the cache-friendly partitioning is not used. Improving thread utilization can indeed improve the performance, especially with a smaller number of threads when the performance is not memory-bound. With cache-friendly partitioning, the difference is fairly small, especially when SIMD is used.

\begin{figure}
  \centering
  \includegraphics[width=0.45\textwidth]{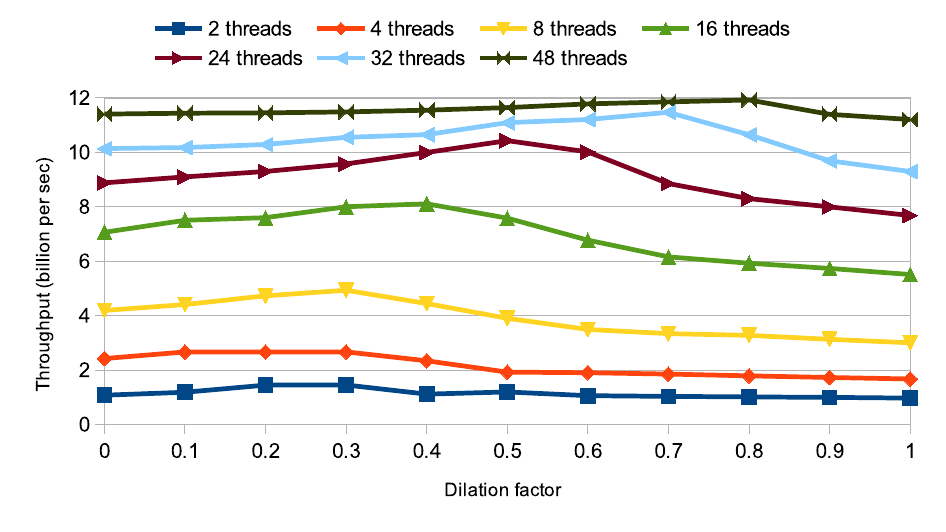}
  \caption{Varying dilation factors with different number of threads}
  \label{fig:exp:dilation_factor}
\end{figure}

As we have demonstrated that the default dilation $d=1$ as used in standard library implementations is suboptimal, users should tune this parameter for better performance. However, it is not very convenient to tune this parameter because although it is the ratio of two different subprocedures, the real performance depends on multiple factors in reality. For example, Figure~\ref{fig:exp:dilation_factor} shows the throughput results with varying dilation factors for the Scalar1 implementation. Using different number of threads, the best dilation factor changes from 0.2 to 0.8, reflecting a changing balance of CPU and memory latencies as memory bandwidth becomes saturated.

From the above results, we observe that if we want to use one more partition (Figures~\ref{fig:multithread3} and \ref{fig:multithread4}), then the dilation factors must be tuned and sometimes it is hard to find a fixed best factor across all configurations. On the other hand, since we have almost the same high performance using the cache-friendly partitioning strategy, without using the extra partition in every iteration, it is more robust to use the partitioning schemes in Figures~\ref{fig:multithread1} and \ref{fig:multithread2}.

\subsubsection{Effect of High-Bandwidth Memory}
\vspace{1.5em}

\begin{figure}
  \centering
  \includegraphics[width=0.45\textwidth]{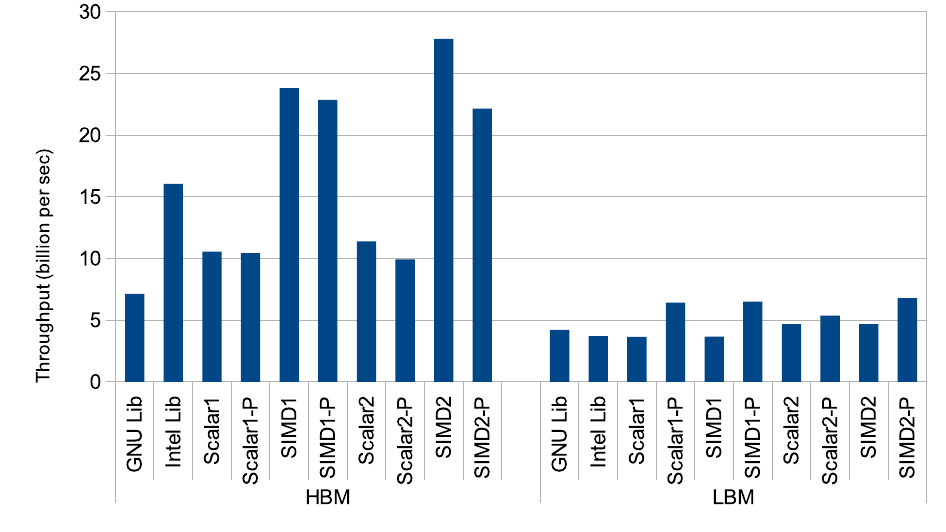}
  \caption{Throughput on Knights Landing}
  \label{fig:exp:phi}
\end{figure}

To study the effect of memory bandwidth, we also conducted experiments on a Xeon Phi machine based on the Knights Landing (KNL) architecture. Although it is not a mainstream machine for database query processing, it has a multi-channel RAM (MCDRAM) providing much higher bandwidth than regular RAM (the load bandwidth is 295 GB per second and the store bandwidth is 220 GB per second). Our experimental machine has 64 physical cores and 16 GB MCDRAM. Figure~\ref{fig:exp:phi} shows the results using 64 threads. It is clear that the cache-friendly partitioning (into L2-cache sized units) helps with every algorithm in regular memory (LBM). In the high bandwidth memory (HBM), we observe a higher throughput for all the methods. However, the partitioning does not improve performance at all, and it is better not to partition with a finer granularity (i.e., the optimal partition size per thread is simply 1/64 of the data size).

To explain these results, observe that there is an overhead to partitioning, including extra synchronization. When data is in RAM, the payoff (reduced memory traffic) is important because the system is memory bound, and so the overhead is worth paying.  When the data is in high-bandwidth memory, the system never becomes memory-bound. With bandwidth as an abundant resource, it does not pay to reduce bandwidth needs by doing extra work.

% \todo{why intel becomes better than gnu?}

% Other experiments:
% \begin{itemize}
% \item effect of autovec
% \item simd width: sse, avx, avx2
% \item in-place, out-of-place
% \item data types: 32-bit ints, 64-bit ints
% \end{itemize}

% Applications: radix sort, stream compaction.

\section{Choosing the Right Algorithm}
\vspace{1.5em}

We now summarize our results, and make recommendations for the efficient implementation of prefix sum algorithms/libraries.

{\bf Observation 1.} It is hard to get configuration parameters like the dilation factor right.  Dilation factors depend on a ratio of subalgorithm speeds that depends on many factors and is hard to calibrate in advance. The performance gains from not wasting one thread are small, particularly if many threads are available, and the risks of suboptimal performance when the wrong dilation factor is used are high.  We should also remark that the partitioning approach also needs a configuration parameter to determine the partition sizes. However, the best partition size appears to be easily determined from the L2 cache size, which can be obtained from the system at runtime.

{\bf Observation 2.} It is worthwhile to pursue bandwidth optimizations like partitioning if bandwidth a bottleneck. Experiments on a machine with high-bandwidth memory show that partitioning helps when the data is in slow RAM, but hurts performance when data is in fast RAM.

{\bf Observation 3.} Over all experiments, the partitioning variant SIMD2-P has the most robust performance in all conditions, while SIMD1-P performed slightly better for in-place computations. Scalar code often reached the same plateau as SIMD code with many threads, but the SIMD code was significantly faster at lower thread counts where bandwidth was not saturated.

{\bf Observation 4.} There are some subtle interactions between in-place/out-of-place choices and algorithm structure, particularly for scalar code. There are also compiler-driven effects, where simpler loop structures in scalar code can be automatically vectorized.

{\bf Observation 5.} Tree-based algorithms are not competitive because of poor memory locality. An alternative vertical SIMD method seems reasonable in theory, but does not perform well because on current machines, gather/scatter instructions are relatively slow.

\section{Conclusions}
\vspace{1.5em}

In this paper we have implemented and compared different ways of computing prefix sums using SIMD and multithreading. We find that efficient SIMD implementations are better able to exploit the sequential access pattern, even if it is not work-efficient in theory. Using AVX-512, we observe more than 3x speedup over scalar code in a single-threaded execution. In a parallel environment, the memory bandwidth quickly becomes the bottleneck and SIMD adds even more pressure on memory access. An efficient multithreaded implementation needs to be cache-friendly, with minimal overhead of thread synchronization. In the experiments we find that the most efficient prefix sum computation using our partitioning technique with better caching is up to 3x faster than standard library implementations that already use SIMD and multithreading.

\balance

% \section{Acknowledgments}

% \newpage

\bibliographystyle{abbrv}
\bibliography{paper}

\begin{thebibliography}{10}

\bibitem{gnulibrary}
Libstdc++ parallel mode.
\newblock \url{https://gcc.gnu.org/onlinedocs/libstdc++/manual/parallel_mode.html}.

\bibitem{intellibrary}
Parallel stl.
\newblock \url{https://github.com/oneapi-src/oneDPL}.

\bibitem{barthels2017distributed}
C.~Barthels, I.~M{\"u}ller, T.~Schneider, G.~Alonso, and T.~Hoefler.
\newblock Distributed join algorithms on thousands of cores.
\newblock {\em Proceedings of the VLDB Endowment}, 10(5):517--528, 2017.

\bibitem{billeter2009efficient}
M.~Billeter, O.~Olsson, and U.~Assarsson.
\newblock Efficient stream compaction on wide simd many-core architectures.
\newblock In {\em Proceedings of the conference on high performance graphics 2009}, pages 159--166, 2009.

\bibitem{blelloch1990vector}
G.~E. Blelloch.
\newblock {\em Vector models for data-parallel computing}, volume~2.
\newblock MIT press Cambridge, 1990.

\bibitem{blellochprefix}
G.~E. Blelloch.
\newblock Prefix sums and their applications.
\newblock {\em Synthesis of Parallel Algorithms}, pages 35--60, 1993.

\bibitem{chaudhuri1992complexity}
S.~Chaudhuri and J.~Radhakrishnan.
\newblock The complexity of parallel prefix problems on small domains.
\newblock In {\em Proceedings., 33rd Annual Symposium on Foundations of Computer Science}, pages 638--647. IEEE, 1992.

\bibitem{cole1989faster}
R.~Cole and U.~Vishkin.
\newblock Faster optimal parallel prefix sums and list ranking.
\newblock {\em Information and computation}, 81(3):334--352, 1989.

\bibitem{fenwick1994new}
P.~M. Fenwick.
\newblock A new data structure for cumulative frequency tables.
\newblock {\em Software: Practice and experience}, 24(3):327--336, 1994.

\bibitem{geffner1999relative}
S.~Geffner, D.~Agrawal, A.~El~Abbadi, and T.~Smith.
\newblock Relative prefix sums: An efficient approach for querying dynamic olap data cubes.
\newblock In {\em Proceedings 15th International Conference on Data Engineering (Cat. No. 99CB36337)}, pages 328--335. IEEE, 1999.

\bibitem{goldberg1995optimal}
T.~Goldberg and U.~Zwick.
\newblock Optimal deterministic approximate parallel prefix sums and their applications.
\newblock In {\em Proceedings Third Israel Symposium on the Theory of Computing and Systems}, pages 220--228. IEEE, 1995.

\bibitem{harris2007parallel}
M.~Harris, S.~Sengupta, and J.~D. Owens.
\newblock Parallel prefix sum (scan) with cuda.
\newblock {\em GPU gems}, 3(39):851--876, 2007.

\bibitem{hetland2019paths}
C.~Hetland, G.~Tziantzioulis, B.~Suchy, M.~Leonard, J.~Han, J.~Albers, N.~Hardavellas, and P.~Dinda.
\newblock Paths to fast barrier synchronization on the node.
\newblock In {\em Proceedings of the 28th International Symposium on High-Performance Parallel and Distributed Computing}, pages 109--120, 2019.

\bibitem{hillis1986data}
W.~D. Hillis and G.~L. Steele~Jr.
\newblock Data parallel algorithms.
\newblock {\em Communications of the ACM}, 29(12):1170--1183, 1986.

\bibitem{ho1997range}
C.-T. Ho, R.~Agrawal, N.~Megiddo, and R.~Srikant.
\newblock Range queries in olap data cubes.
\newblock {\em ACM SIGMOD Record}, 26(2):73--88, 1997.

\bibitem{kim2009sort}
C.~Kim, T.~Kaldewey, V.~W. Lee, E.~Sedlar, A.~D. Nguyen, N.~Satish, J.~Chhugani, A.~Di~Blas, and P.~Dubey.
\newblock Sort vs. hash revisited: Fast join implementation on modern multi-core cpus.
\newblock {\em Proceedings of the VLDB Endowment}, 2(2):1378--1389, 2009.

\bibitem{klemm2012extending}
M.~Klemm, A.~Duran, X.~Tian, H.~Saito, D.~Caballero, and X.~Martorell.
\newblock Extending openmp* with vector constructs for modern multicore simd architectures.
\newblock In {\em International Workshop on OpenMP}, pages 59--72. Springer, 2012.

\bibitem{kohlhoff2012k}
K.~J. Kohlhoff, V.~S. Pande, and R.~B. Altman.
\newblock K-means for parallel architectures using all-prefix-sum sorting and updating steps.
\newblock {\em IEEE Transactions on Parallel and Distributed Systems}, 24(8):1602--1612, 2012.

\bibitem{ladner1980parallel}
R.~E. Ladner and M.~J. Fischer.
\newblock Parallel prefix computation.
\newblock {\em Journal of the ACM (JACM)}, 27(4):831--838, 1980.

\bibitem{leischner2010gpu}
N.~Leischner, V.~Osipov, and P.~Sanders.
\newblock Gpu sample sort.
\newblock In {\em 2010 IEEE International Symposium on Parallel \& Distributed Processing (IPDPS)}, pages 1--10. IEEE, 2010.

\bibitem{lemire2002wavelet}
D.~Lemire.
\newblock Wavelet-based relative prefix sum methods for range sum queries in data cubes.
\newblock In {\em Proceedings of the 2002 conference of the Centre for Advanced Studies on Collaborative research}, page~6. IBM Press, 2002.

\bibitem{lemire2015decoding}
D.~Lemire and L.~Boytsov.
\newblock Decoding billions of integers per second through vectorization.
\newblock {\em Software: Practice and Experience}, 45(1):1--29, 2015.

\bibitem{lemire2016simd}
D.~Lemire, L.~Boytsov, and N.~Kurz.
\newblock Simd compression and the intersection of sorted integers.
\newblock {\em Software: Practice and Experience}, 46(6):723--749, 2016.

\bibitem{maleki2011evaluation}
S.~Maleki, Y.~Gao, M.~J. Garzar, T.~Wong, D.~A. Padua, et~al.
\newblock An evaluation of vectorizing compilers.
\newblock In {\em 2011 International Conference on Parallel Architectures and Compilation Techniques}, pages 372--382. IEEE, 2011.

\bibitem{polychroniou2015rethinking}
O.~Polychroniou, A.~Raghavan, and K.~A. Ross.
\newblock Rethinking simd vectorization for in-memory databases.
\newblock In {\em Proceedings of the 2015 ACM SIGMOD International Conference on Management of Data}, pages 1493--1508, 2015.

\bibitem{raman2007succinct}
R.~Raman, V.~Raman, and S.~R. Satti.
\newblock Succinct indexable dictionaries with applications to encoding k-ary trees, prefix sums and multisets.
\newblock {\em ACM Transactions on Algorithms (TALG)}, 3(4):43--es, 2007.

\bibitem{rodchenko2015effective}
A.~Rodchenko, A.~Nisbet, A.~Pop, and M.~Luj{\'a}n.
\newblock Effective barrier synchronization on intel xeon phi coprocessor.
\newblock In {\em European Conference on Parallel Processing}, pages 588--600. Springer, 2015.

\bibitem{sanders2006parallel}
P.~Sanders and J.~L. Tr{\"a}ff.
\newblock Parallel prefix (scan) algorithms for mpi.
\newblock In {\em European Parallel Virtual Machine/Message Passing Interface Users’ Group Meeting}, pages 49--57. Springer, 2006.

\bibitem{satish2009designing}
N.~Satish, M.~Harris, and M.~Garland.
\newblock Designing efficient sorting algorithms for manycore gpus.
\newblock In {\em 2009 IEEE International Symposium on Parallel \& Distributed Processing}, pages 1--10. IEEE, 2009.

\bibitem{satish2010fast}
N.~Satish, C.~Kim, J.~Chhugani, A.~D. Nguyen, V.~W. Lee, D.~Kim, and P.~Dubey.
\newblock Fast sort on cpus and gpus: a case for bandwidth oblivious simd sort.
\newblock In {\em Proceedings of the 2010 ACM SIGMOD International Conference on Management of data}, pages 351--362, 2010.

\bibitem{sengupta2007scan}
S.~Sengupta, M.~Harris, Y.~Zhang, and J.~D. Owens.
\newblock Scan primitives for gpu computing.
\newblock 2007.

\bibitem{sengupta2006work}
S.~Sengupta, A.~Lefohn, and J.~D. Owens.
\newblock A work-efficient step-efficient prefix sum algorithm.
\newblock 2006.

\bibitem{singler2008gnu}
J.~Singler and B.~Konsik.
\newblock The gnu libstdc++ parallel mode: software engineering considerations.
\newblock In {\em Proceedings of the 1st international workshop on Multicore software engineering}, pages 15--22, 2008.

\bibitem{singler2007mcstl}
J.~Singler, P.~Sanders, and F.~Putze.
\newblock Mcstl: The multi-core standard template library.
\newblock In {\em European Conference on Parallel Processing}, pages 682--694. Springer, 2007.

\bibitem{yan2013streamscan}
S.~Yan, G.~Long, and Y.~Zhang.
\newblock Streamscan: fast scan algorithms for gpus without global barrier synchronization.
\newblock In {\em Proceedings of the 18th ACM SIGPLAN symposium on Principles and practice of parallel programming}, pages 229--238, 2013.

\end{thebibliography}

% \begin{appendix}
% \end{appendix}

\end{document}